\documentclass[aps,showpacs,twocolumn,floatfix,pra,superscriptaddress]{revtex4-1}
\usepackage{graphicx}
\usepackage{amsfonts}
\usepackage{amsmath}
\usepackage{amssymb}
\usepackage[usenames,dvipsnames]{color}

\begin{document}
\title{Space-time velocity correlation function for random walks}
\author{V. Zaburdaev}
\affiliation{Max Planck Institute for the Physics of Complex Systems, N\"{o}thnitzer Str. 38, D-01187 Dresden, Germany}
\author{S. Denisov}
\author{P. H\"{a}nggi}
\affiliation{Institute of Physics, University of Augsburg, Universit\"{a}tstr.~1, D-86159 Augsburg, Germany}
\begin{abstract}
Space-time correlation functions constitute a useful instrument from the research toolkit of continuous-media and many-body physics.
We adopt here this concept for single-particle random walks and demonstrate that the corresponding space-time velocity auto-correlation functions
reveal  correlations which extend in time much longer than estimated with the commonly employed temporal
correlation functions. A generic feature of considered random-walk processes is an effect of velocity echo identified by
the existence of time-dependent regions where most of the walkers are moving in the direction opposite to their initial motion.
We discuss the relevance of the space-time velocity correlation functions for the experimental studies of cold atom dynamics in
an optical potential and  charge transport on micro- and nano-scales.

\end{abstract}
\maketitle

{\em Introduction}.
Finiteness of velocity is a fundamental property of any physical process  taking place in space and time.
This concept was  incorporated  in the framework of random-walk theory \cite{pearson}, a formalism that is
particularly successful in describing diffusion phenomena \cite{rw2}. If the speed of a walking particle is constant,
the coupling between the distance traveled and the time it takes leads to the confinement of the spreading process to a casual cone.
Within the cone  the density of particles
is described by the phenomenological diffusion equation \cite{keller}.  The space-time coupling  also regularizes fast super-diffusion
by removing possibly unphysical divergences from the momenta of the corresponding processes \cite{yossi1, yossi2}.

The velocity of a random-walk process can be treated as an additional dynamical variable whose evolution is itself
a random process. The Green-Kubo relation \cite{kubo}
highlights the importance of the corresponding temporal auto-correlation function by connecting its integral to the diffusion coefficient of
the process  \cite{taylor}. It is evident, however, that the velocity auto-correlations of a random walk cannot last longer than the time
between two consecutive re-orientation events.

Following this premise, we next ask whether more extended correlations  can be detected by unfolding the velocity correlation function into the \textit{spatial} domain.
Here, we answer this question positively by introducing a characteristic that reveals hitherto unnoticed properties of random-walk processes. We define the space-time (\textit{s-t}) velocity auto-correlation function
for single-particle random walks by adopting a concept widely used in fluid dynamics \cite{fluid}, gas and plasma kinetics \cite{kinetic}.
For models yielding normal and super-diffusive dynamics, we show that this function helps to uncover long-lived
correlations that  extend beyond the horizon dictated by the standard temporal correlation function.
Furthermore, the unfolding into the spatial domain allows for a meaningful description of velocity correlations when the temporal correlation
function simply does not exist. We argue that the process-specific generalized correlation function can be accessed experimentally.
Thus it can serve as a tool to determine stochastic processes underlying  macroscopic diffusion phenomena observed in experiments.

{\em Continuous-time random walks and \textit{s-t} velocity auto-correlation function}.
We consider single-particle processes  that belong to a class of  continuous-time random walks (CTRWs) \cite{rw2}.
In its simplest one-dimensional realization, such a walk is performed by a particle moving ballistically  with a fixed velocity $v_i$ between two turning events.
The duration of the $i$-th 'flight', that is the time interval between two consecutive turnings, $\tau_i = t_{i+1} - t_i$, is governed by a probability density
function (PDF) $\psi(\tau)$. At the end of each flight the particle changes its velocity to a new random value $v_{i+1}$,
 sampled from the PDF $h(v)$, and then starts a new flight.
Two random variables, $\tau_i$ and $v_i$, are statistically independent but the spatial and temporal evolution of the walker during the flight is coupled,  $x_{i+1} - x_i = v_i\tau_i$.
This general setup is able to reproduce normal and anomalous diffusion regimes \cite{zabu1, zabu2}, and there
is a multitude of real-life systems and processes  whose dynamics can be described by this model \cite{solomon, atoms, foraging, mussels, chemo}.

The key property of the described CTRW model is a well-defined velocity of a walker at any
instant of time. It allows us to introduce space-time velocity auto-correlation function for a single-particle process
by redefining the conventional expression  \cite{fluid, kinetic}
\begin{equation}
\mathcal{C}_{vv}(x,t)=\left<v(0,0) v(x,t)\right> .
\label{cor_def}
\end{equation}

That is, we assume that the particle starts  its  walk with initial velocity $v(x=0,t=0) = v_0$.
After a time $t$ the particle is found at the point $x$ with some velocity
$v(x,t)$. 
To estimate $\mathcal{C}_{vv}(x,t)$, an observer at time $t$ averages the product of the actual and the initial
velocities of all particles that are located within a bin  $[x, x+dx]$.
The so-measured quantity can be formalized as
\begin{equation}
\mathcal{C}_{vv}(x,t) = \int\limits_{-\infty}^{\infty}\int\limits_{-\infty}^{\infty} v v_0 \frac{P(v,x,v_0,t)}{P(x,t)}dv_0dv,
\label{cor_form}
\end{equation}
where $P(v,x,v_0,t)$ is the joint PDF for a particle to start with velocity $v_0$ and to be in the point $x$ at time $t$ with velocity $v$ \cite{pdfexplained}.
Since the particle has first to arrive to the point $x$ for the measurement to occur, we use the formula for the conditional probability
and divide the joint
density by the spatial PDF $P(x,t)$. The latter is a well-studied characteristic of the CTRW processes \cite{rw2}.
In contrast, a challenging quantity to tackle is the joint probability of particles' positions and velocities.
To focus on its role, we introduce the spatial density of the velocity correlation function,
\begin{equation}
C(x,t)=\int\limits_{-\infty}^{\infty}\int\limits_{-\infty}^{\infty} v v_0 P(v,x,t|v_0)h(v_0)dv_0dv.
\label{vacf_integral}
\end{equation}
Here we split the joint PDF,  $P(v,x,v_0,t)=P(x,v,t |v_0)h(v_0)$, in order to factorize the averaging
with respect to initial velocities \cite{dimension}.
There are two noteworthy features of this new quantity.
First, after the integration over $x$, Eq.~(\ref{vacf_integral}) yields the standard temporal
velocity auto-correlation function $C(t)=\left<v(0)v(t)\right>$.
Second, to return to the original {\em s-t} velocity auto-correlation function, Eq.~(\ref{cor_def}),
$C(x,t)$ has to be normalized with the spatial density $P(x,t)$,
\begin{equation}
\mathcal{C}_{vv}(x,t)=C(x,t)/P(x,t).
\label{connection}
\end{equation}
Therefore, our further analysis is restricted to the function $C(x,t)$,
while the obtained results can be immediately mapped onto $\mathcal{C}_{vv}(x,t)$ by virtue of Eq.~(\ref{connection}).

We are now set to derive an equation for $P(v,x,t|v_0)$.
We first introduce the frequency of velocity changes, $\nu_{v_0}(x,t)$, with $\nu_{v_0}(x,t)dxdt$ counting the number of particles whose flights ended $[x, x+dx]$ during the
time interval $[t, t+dt]$.
The additional subscript $v_0$ tracks the history of particles and denotes only those which had velocity $v_0$ at $t=0$.
The balance equation for $\nu_{v_0}$ assumes the form of an integral equation, reading
\begin{eqnarray}
\nu_{v_0}(x,t)=\int\limits_{-\infty}^{+\infty}&dv&\int\limits_{0}^{t}\nu_{v_0}(x-v\tau,t-\tau)h(v)\psi(\tau)d\tau\nonumber\\
&+&\psi(t)\delta(x-v_0t).
\label{freq_cor}
\end{eqnarray}
A particle changes its velocity at the end of the flight of duration $\tau$ that was initiated at the point $x-v\tau$.
Multiplication by $\psi(\tau)d\tau$ and $h(v)dv$ yields the probability of having a flight-time $\tau$ and velocity $v$.
We also assume that all particles start their random walk at $t=0$ and $x=0$. The last term on the right hand side accounts for the particles
that finish their very first flight at the  given moment of time $t$.
Correspondingly, $P(v,x,t|v_0)$ is:
\begin{eqnarray}
P(v,x,t|v_0)=& &\int\limits_{0}^{t}\nu_{v_0}(x-v\tau,t-\tau)h(v)\Psi(\tau)d\tau\nonumber\\&+&\Psi(t)\delta(x-v_{0}t)\delta(v-v_0).
\label{density_cor}
\end{eqnarray}
A particle has a velocity $v$ at  point $(x,t)$ if it has previously changed its velocity at time $t-\tau$ and still is in the process of flight with velocity $v$.
The probability to stay in the flight until time $t$ is given by $\Psi(t)=1-\int_{0}^{t}\psi(\tau)d\tau$. The second term on the right hand side of Eq. (\ref{density_cor})
accounts for the particles that are still in their first flight. 
The above two equations
can be resolved by using  a combined Fourier-Laplace transform with respect to $x$ and $t$
which turns the convolution-type integrals into algebraic products. We use $\mathcal{L}[...]$ and a hat to denote the Laplace and Fourier transforms,  and a tilde for a combination of the two, whereas $k$ and $s$ denote coordinates in the
Fourier and Laplace spaces. We find $\tilde{P}(v,v_0,k,s)$, and by using Eq.(\ref{vacf_integral}), obtain
the general expression for the  velocity correlation density in the Fourier-Laplace space,
\begin{widetext}
\begin{equation}
\widetilde{C}(k,s)=\frac{\mathcal{L}\left[\int\limits_{-\infty}^{\infty}\Psi(\tau)e^{-ikv\tau}vh(v)dv\right]~~\mathcal{L}\left[\int\limits_{-\infty}^{\infty}\psi(\tau)e^{-ikv_0\tau}v_0h(v_0)dv_0\right]}{1-\mathcal{L}\left[\widehat{h}(k\tau)\psi(\tau)\right]}+\mathcal{L}\left[\int\limits_{-\infty}^{\infty}\Psi(t)e^{-ikv_0\tau}v_{0}^{2}h(v_0)dv_0\right].
\label{cks}
\end{equation}
\end{widetext}
Equation (\ref{cks}) constitutes the central result of this Letter.

\begin{figure}[t]
\center
\includegraphics[width=0.45\textwidth]{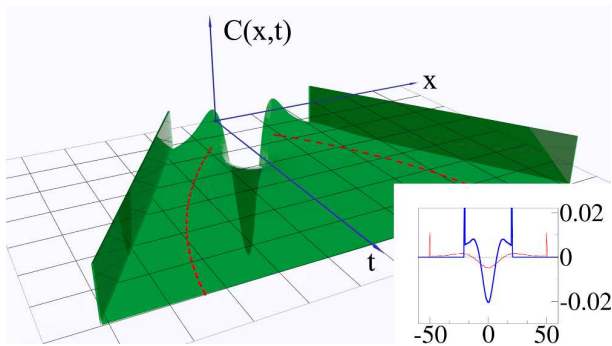}
\caption[Models]
{(color online) Density of the space-time velocity correlation function  for the super-diffusive  L\'{e}vy walk, Eq.~(\ref{csuperdif}),
with $\gamma=3/2$ as a function of $x$ and $t$. The red dashed lines indicate the positions of local maxima $x^\pm_{\text{m}}$ on the
$x - t$ plane which follow the power-law scaling $x^\pm_{\text{m}}\propto \pm t^{1/\gamma}$, while  the height of the maxima decays as
$t^{-1-1/\gamma}$. The inset depicts spatial profiles of $C(x,t)$ for two different instants of time, $t = 20$ (heavy blue line)  and $50$ (light red line).}
\label{fig1}
\end{figure}
When integrating this function over $x$, the first term, describing the contribution from the particles
that have changed their velocities several times,
vanishes.  It is only the particles remaining in their first flight
that contribute to $C(t)$. This implies that  $C(t)=\left<v^2\right>\Psi(t)$.
Below we consider all regimes of diffusion which are possible in the current random walk model, ranging from the
standard diffusion to ballistic super-diffusion.

{\em L\'{e}vy walks: from normal to ballistic super-diffusion.}
The L\'{e}vy walk process \cite{rw2} is a particular case with a bimodal choice of the velocity PDF, $h(v)=\left[\delta(v-u_0)+\delta(v+u_0)\right]/2$, and
the PDF of flight-time
\begin{equation}
\psi(\tau)=\frac{\gamma}{\tau_0}\frac{1}{(1+\tau/\tau_0)^{1+\gamma}}\;,
\label{psi1}
\end{equation}
where $\tau_0$ sets the time-scale of the process. The positive scaling exponent  $\gamma > 0$ plays a key role in defining the  type of the diffusion.
If  $\gamma > 2$ then the mean square of the flight-time is finite and the process reproduces normal, Brownian-like diffusion, with the linear scaling of the mean squared displacement,
$\langle x^2(t)\rangle \propto t$ \cite{rw2}. The mean squared flight-time  diverges for $1<\gamma<2$, which leads
to anomalously fast spreading, $\langle x^2(t)\rangle \propto t^{3 - \gamma}$ \cite{yossi1, geisel}.
In the extreme case of $0<\gamma<1$, the anomaly is strong and prevails the evolution, because even the first moment of the flight-time diverges;
so that the mean squared displacement scales ballistically, $\langle x^2(t)\rangle \propto t^{2}$.
In all cases, however, the density of particles $P(x,t)$ is confined to the ballistic cone, $ x \in [-u_0 t, u_0 t]$.

When $\gamma>2$,  inside of the casual cone and in the asymptotic limit,  the density of particles obeys the standard diffusion equation \cite{keller},
\begin{equation}
\frac{\partial P(x,t)}{\partial t}=D\triangle P(x,t); \quad D=\frac{u_0^2\tau_0}{\gamma-2}.
\label{diff}
\end{equation}
In this case, the first term of equation (\ref{cks}) in normal coordinates  reduces to a  simple result:
\begin{equation}
C_{\text{centr}}(x,t)=\frac{u_0^2D\tau_0}{\gamma-1}\triangle P(x,t)=\frac{u_0^2\tau_0}{\gamma-1}\frac{\partial P(x,t)}{\partial t}.
\label{cor2}
\end{equation}
It reveals an interesting relation between the PDF of the process and the corresponding  correlation density function,
namely that $C(x,t)$ is proportional to  $\partial P(x,t)/\partial t$ and, according to Eq. (\ref{connection}),
the normalized {\em s-t} correlation function $\mathcal{C}_{vv}(x,t)\propto \partial \ln P(x,t)/\partial t$.
Note that the contribution of the second term in Eq.~(\ref{cks})
corresponds to the ballistic delta-peaks, running with the speed $u_0$ and decaying in time according to
$\Psi(t)$.  Ballistic peaks are the hallmark  of  L\'{e}vy walks \cite{yossi3}.
Their contribution is typically considered to be asymptotically vanishing in the
regime of standard  diffusion. However, only these peaks contribute to the temporal correlation function $C(t)$ and therefore cannot be neglected.
By taking into account that the number of particles in the peaks also decays as $\Psi(t)$,
it immediately follows from  Eq.~(\ref{connection}) that the normalized {\em s-t} correlation function, Eq.~(\ref{cor_def}),
remains constant at the ballistic fronts, $\mathcal{C}_{vv}(x=\pm u_0t,t)=u_0^2$.

The results presented by Eqs.~(\ref{diff}, \ref{cor2})  are valid for an arbitrary choice of $\psi(t)$ that has finite second
moment, $\int_0^\infty \tau^2 \psi(\tau) d\tau < \infty$, including
the case of the  exponential PDF $\psi(\tau) = e^{-\tau/\tau_0}/\tau_0$.
In this case  $C(t)=u_0^2e^{-t/\tau_0}$ while $C(x,t)$, for example,  at $x=0$, scales like $t^{-3/2}$. 
This example highlights the fact that the $s-t$ velocity correlation functions provide access to long-lived correlations and
therefore increase the chance of their detection.

For $1 < \gamma < 2$ the mean squared flight-time  diverges.
It induces a super-diffusive behavior with the density of particles obeying a generalized diffusion equation  \cite{zaslik},
\begin{equation}
\frac{\partial P(x,t)}{\partial t}=-K(-\triangle)^{\gamma/2} P(x,t);
\label{superdiff}
\end{equation}
where $\quad K=\tau_0^{\gamma-1}u_0^{\gamma}(\gamma-1)\Gamma[1-\gamma]\cos(\pi\gamma/2)$ and $\triangle^{\gamma/2}$ is the fractional Laplacian operator \cite{laplacian}.
Note that this description is valid in the inner part of the casual cone only.
In there $C(x,t)$ is proportional to the fractional Laplacian of the density of particles,
\begin{equation}
C_{\text{centr}}(x,t)=-\frac{u_0^2 K\tau_0}{\gamma-1}(-\triangle)^{\gamma/2}P(x,t)=\frac{u_0^2\tau_0}{\gamma-1}\frac{\partial P(x,t)}{\partial t};
\label{csuperdif}
\end{equation}
or,  by virtue of the Eq. (\ref{superdiff}),  to the
time derivative of this density. Therefore, the velocity auto-correlations are negative near  the point $x=0$, see in Fig. 1.
Upon the departure from the origin the
correlation density becomes positive and produces two local maxima.

By setting $\gamma < 1$ in Eq.~(\ref{psi1}), one can enhance  the anomalous character of the process. The  average flight-time of particles diverges and this implies a ballistic scaling for the
density of particles. Agian,  $C(x,t)$  can be evaluated in the Fourier-Laplace space.
As an illustration, we consider the case $\gamma=1/2$ where both quantities,  $P(x,t)$ and $C(x,t)$, can be expressed in terms of analytic functions \cite{zabu1, zabu2}.
The density exhibits an $U$-shaped profile, diverging at the ballistic fronts: $P(x,t)=\theta(u_0t-|x|)/[\pi(t^2u_{0}^{2}-x^2)^{1/2}]$.
The correlation density function behaves similarly:
\begin{eqnarray}
C(x,t) &=&\label{cballistic}\\
&=&u_{0}^{2}\tau_0^{1/2}\left[\frac{\delta(x-u_0t)}{t^{1/2}}+\frac{\delta(x+u_0t)}{t^{1/2}}-\frac{\theta(u_0t-|x|)}{2t^{3/2}u_0}\right]\nonumber
\end{eqnarray}
From this  result it follows  that  velocity correlations are negative and nearly constant inside the ballistic cone.
The relative decay rate $P(x,t)/C(x,t)$  is  of the order $t^{1/2}$ now.
The profiles evaluated via  direct numerical simulations of the random walk, cf. Fig.~2(a),  perfectly match  the analytical prediction.

{\em Velocity-induced super-diffusion.}
Regimes of diffusion described by the L\'{e}vy walk model are bound from above by the ballistic propagation so that  no particles can cross the  front $|x|=u_0t$.
One possible way to overcome this limitation is to allow the flight speed  to have a broad distribution.
Assume that $h(v)$ is a  Lorentz distribution, $h(v)=(\pi u_0)^{-1}(1+v^2/u_0^2)^{-1}$,
a velocity PDF frequently employed in plasma
 and kinetic theories \cite{plasma}.
For such velocity PDF, the density of particles is independent of the flight-time distribution and also
possesses the Lorentz shape in the asymptotic limit \cite{zabu2}: $P(x,t)=u_0t/[\pi(u_0^2t^2+x^2)]$.
The expression for the  correlation density function then acquires the form:
\begin{equation}
C(x,t)=u_0^2\left[-\frac{u_0t}{\pi(u_0^2t^2+x^2)}+\frac{1}{u_0t}\Psi(t)\right]. \label{Canalyt}
\end{equation}
The first term is proportional to the density but  with the opposite sign and the second depends on the flight-time distribution.
Note that the integral of $C(x,t)$ with respect to $x$ diverges when $\Psi(t)\neq 0$, meaning an infinite $C(t)$.
In clear contrast,  density of the velocity correlation function is well-defined, see Fig. 2(b).
\begin{figure}[t]
\center
\includegraphics[width=0.45\textwidth]{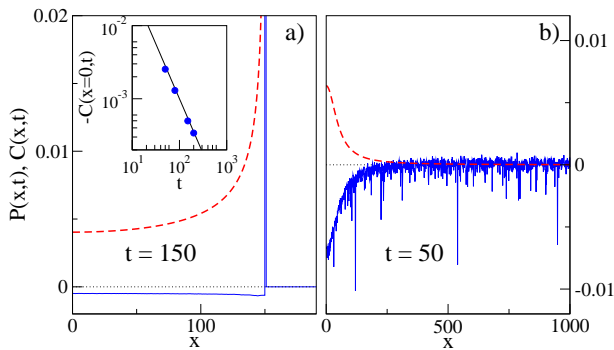}
\caption[Models]
{(color online) Densities of particles, $P(x,t)$ (dashed red lines), and the spatial
density of  velocity correlations, $C(x,t)$ (solid  blue lines) for
(a) the L\'{e}vy walk in the regime of ballistic super-diffusion ($\gamma=1/2$), Eq. (\ref{cballistic}),
 and (b) the case of Lorentzian velocity PDF, Eq.(\ref{Canalyt}), with $\psi(\tau) = \delta(\tau - 1)$.
Correlation functions were obtained  by averaging over $10^8$ realizations of the corresponding process.
The inset shows $C(0,t)$ at four different instants of time. Solid line corresponds to the power-law $t^{-3/2}$.
All other parameters are set to one.
}
\label{fig2}
\end{figure}

{\em Discussion.} In all considered regimes there is a region of negative correlations at the vicinity of the starting point.
This means that  majority of particles found there are flying in the direction opposite to that of their initial motion, which we call an {\em echo effect}.
The shape of the echo region, the time-scaling of its width and height are model-specific
characteristics. Simulations of a stochastic process described by a system of Langevin equations \cite{sup} show analogous results,
which suggests
that our  findings are applicable to a broad class of stochastic transport processes characterized
by finite velocity of moving particles.

Perhaps the best candidate for the analog simulation of super-diffusive continuous-time random walks is a
cold atom moving in a periodic optical potential.
There are strong evidences, both theoretical \cite{atoms}  and experimental \cite{katori, davidson},
that the diffusion of the atom along the potential is anomalous and can be reproduced with L\'{e}vy-walk models \cite{davidson, barkai}.
We suggest that the velocity correlation function can be measured in  experiments and thus will help to build a proper
microscopic model. It is possible to prepare a strongly localized ensemble with all atoms having near equal velocities, either by
sudden release of atoms from a ballistically moving deep optical well or by using more exotic setups  \cite{laser, soliton}.
For such initial states, the measurement of
of $C(x,t)$  is equivalent to finding of the PDF $P(v,x,t)$, see Eq. (\ref{vacf_integral}). The PDF of instantaneous velocities, $P(v,t)$,
can be measured with the routine time-of-flight
technique \cite{tof}, when velocity of an atom is transformed into the atom position, which is then recorded by using the florescence effect.
It is also feasible to measure the spatial distribution of atoms, $P(x,t)$, by using the florescence image of the cloud \cite{tof}.
A measurement of the space-dependent velocity PDF  requires the implementation of the time-of-flight technique combined with a
consecutive de-convolution procedure \cite{convolution}.
By knowing the spatial distribution of the cloud at time $t$ (obtained from another experiment under the same conditions),
the de-convolution transform can
be performed numerically on the fluorescent snapshot of the time-of-flight experiment to reconstruct $P(x,v,t)$ and,
consequently to calculate $C(x,t)$. More sophisticated measurement protocols can be also developed \cite{duan}.

The negative velocity echo, being a distinctive footprint of CTRWs, can be used
as a benchmark to judge on the validity of random-walk approaches to the charge transport on nanoscale.
The velocity echo can be detected by measuring the  current-current \textit{s-t} correlations after a local injection of
electrons into a nanotube \cite{tip}. A recently developed  terahertz time-domain measurement technique
\cite{readout} can be used for the readout. This noninvasive method is capable to resolve  the electron  dynamics on
picosecond timescale thus providing an insight into the real-time propagation of electrons along the nanotube \cite{readout1}.
Another type of systems, where short injection pulses of charge carriers are routinely used to probe charge transport, are slabs
of  semiconductors \cite{bulk}. It is noteworthy that recent experiments have revealed a good agreement between the
dynamics of holes in a bulk of $n$-doped $InP$ slab  and a L\'{e}vy-walk model \cite{bulk1}.

{\em Conclusions.}
The spatial dependence of the {\em s-t} density of velocity auto-correlations, Eq. (\ref{vacf_integral}),  can be decomposed into two contributions.
The first  is produced by the particles which have performed several flights before the observation time.
The second originates from the particles that are still in their first flight. For any random walk with finite average flight-times, the central
part of the velocity correlation pattern can be calculated as the time derivative of the particle's density.

We  believe that the concept developed here can be utilized for any process that can be described as a continuous-time random walk,
where finite velocities can be attributed to the diffusing entities at any instant of time \cite{politi, dhar, barkai3}.
It is possible to generalize the spatiotemporal velocity auto-correlation functions  to the case of two-dimensional random walks \cite{sup},
thus reaching another level of detail in the analysis of the complex transport phenomena \cite{solomon, foraging, mussels, chemo, katori,
our, politi2, barkai4}.

This work has been supported by the DFG Grant HA1517/31-2 (S.D. and P.H.).

\begin{center}
\textbf{Supplementary Material}
\end{center}

\section{A case of Langevin dynamics}

The Langevin equation (LE) is another fundamental model, complementary to random walks,  used for microscopic description of diffusive transport. 
A close relationship between the LE and random walk models suggests that the results obtained for the last should stay 
valid for the former (and vice versa). It is natural, therefore, to test the concept of the space-time velocity correlation 
function on the standard one-dimensional Langevin process,  which is governed by the following pair of linear differential equations:
\begin{eqnarray}
&\dot{x}&=v\\
&\dot{v}&=-\gamma v+\xi(t).
\end{eqnarray}
Here $\xi(t)$ is white Gaussian noise and $\left<\xi(t)\xi(t')\right>=D_v\delta(t-t')$. 
Similar to the considered random walks, the key feature of the LE process is a finite velocity 
of a diffusing particle at each moment of time. In the asymptotic regime, the spatial PDF of the particles obeys the standard diffusion 
equation with $D=D_v/(2\gamma^2)$. There is also no problem with derivation of the corresponding velocity auto-correlation function, $C(t)$,  
see Ref. \cite{OU}.
However, we were unable to obtain analytic expression for the {\em s-t} correlation function, $C(x,t)$; its derivation  
for the case of LE remains an open problem worth of further investigation \cite{eval}.

In Fig. 1 we present the results of numerical simulations obtained by propagating equations (1-2) in time.
As expected, the calculated space-time velocity correlation function fits with the analytic prediction, see Eq.(6) in the main text.
Namely, the function $C(x,t)$ is given by the first time derivative of the  spatial 
PDF, obtained from the corresponding diffusion equation. 

\begin{figure}[t]
\center
\includegraphics[width=0.35\textwidth]{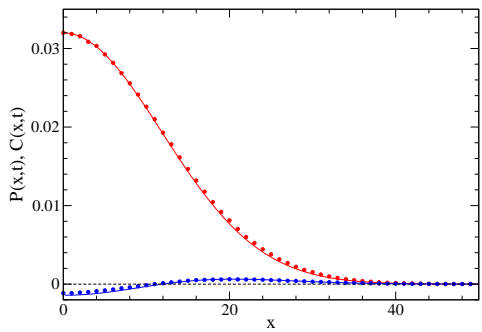}
\caption[LE]
{Spatial density $P(x,t)$ (red dots) and space-time velocity auto-correlation function $C(r,t)$ (blue dots) obtained by sampling 
evolution of the system (1 - 2) for $t = 40$. The parameters are $\gamma = 0.1$ and $D = 0.05$. The blue line corresponds to  the PDF
obtained from the corresponding diffusion equation, the red line is given by the time derivative of the PDF, see Eq. (6) in the main text.}
\label{fig2}
\end{figure}

\section{Generalizations to higher dimensions}

The concept of the \textit{s-t} velocity correlation function can be extended to higher dimensions in a straightforward manner.

In the case when velocity directions at each step of a random walk are isotropic and independent random variables, general expressions
retain their form with the only modification that the corresponding  velocity and coordinates, see Eqs.(3-5) in the main text, have to be replaced by the corresponding vector quantities, $x,k,v\rightarrow \mathbf{r},\mathbf{k},\mathbf{v}$. 
The increase of the dimensionality has certain consequences though; 
even for the regime of standard diffusion the  asymptotic analysis is  more cumbersome. However, this does not affect the qualitative outcome. 
Fig.2 depicts the results of simulations together with analytical expressions  for a two-dimensional random walk
in the regime of normal diffusion. As its one-dimensional predecessor,  the velocity correlation function is given here 
by the time derivative of the radial particle PDF $P(r,t)$.

Two- and three-dimensional random walks are key ingredient of the random coil 
model,  simple yet  powerful concept popular in polymer physics \cite{WK, FL}. 
The random coil model assumes that each monomer --which is a step of a random walk-- 
has a length $l$ and is randomly oriented in space, see the inset on Fig. 2. 
Therefore each configuration of the polymer corresponds to some realization of a random 
walk of a fixed step length $l$ with the directions of consequent steps being uncorrelated. 
The length of the polymer $L=n\cdot l$ corresponds to the duration of the corresponding random walk process. 
Remarkably, our results allow to find correlations between the directions of monomers which 
are separated by a distance $s$ along the backbone of the polymer (see green shaded path on the 
inset of Fig. 2) while being separated in real space by a distance $|\mathbf{r}|$. 
This information might be useful for the estimation of monomer-monomer interactions whose strength
depends on the alignment of monomers; one example is the homologous recombination 
of chromosomes \cite{PH} during meiosis.
 
\begin{figure}[b]
\center
\includegraphics[width=0.45\textwidth]{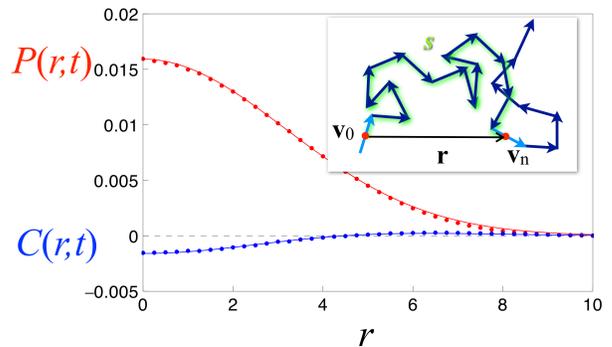}
\caption[Models]
{Radial density $P(r,t)$ (red line) and space-time velocity auto-correlation functions $C(r,t)$ (blue line) 
for the two-dimensional diffusion. Dots show the results on numerical simulations. 
Speed and average flight time in the exponential distribution are set to unity, $t=10$. }
\label{fig2}
\end{figure}


\begin{thebibliography}{99}
\bibitem{pdfexplained} PDF $P(v,x,v_0,t)$ has a meaning of density with respect to $x,v$ and $v_0$ whereas $t$ is a parameter. More formally it could be written as $P(v,x,v_0|x_0,t_0; t)$, however we always set $x_0=0$ and $t_0=0$ and in the text use a simplified notation for the joint PDF.
\bibitem{dimension} Although having meaning of the spatial density, $C(x,t)$ is allowed to be negative.
It has a dimension $[\text{m}/\text{s}^2]$.
\end{thebibliography}
\begin{thebibliography}{99}
\bibitem{eval} This problem, in principle, can be approached from the opposite end, by using 
the Fokker-Planck equation. This equation governs the evolution of the PDF $P(x,v,t)$, which process is parametrized 
by the initial conditions including $v_0$. The averaging of the product $v(x,t)\cdot v_0$ with the PDF $P(x,v,t)$, weighted by 
the Maxwellian distribution $h(v_0)$, will yield $C(x,t)$.
\end{thebibliography}
\end{document}